\documentclass[floatfix,aps,prd,showpacs,amsmath,nofootinbib,preprintnumbers,twocolumn]{revtex4}

\usepackage{graphicx}
\usepackage{bm}

\newcommand{\figurewidth}{2.7in}

\def\half{{1\over 2}}

\def\p{\partial}

\def\half{{1\over 2}}
\def\({\left(}
\def\){\right)}
\def\[{\left[}
\def\]{\right]}

\def\e{\begin{equation}}
\def\q{\end{equation}}
\def\m{\begin{eqnarray}}
\def\n{\end{eqnarray}}

\begin{document}

\title{A polynomial $f(R )$ inflation model}

\author{Qing-Guo Huang}\email{huangqg@itp.ac.cn}
\affiliation{State Key Laboratory of Theoretical Physics, Institute of Theoretical Physics, 
Chinese Academy of Science, Beijing 100190, People's Republic of China}

\date{\today}

\begin{abstract}

Motivated by the ultraviolet complete theory of quantum gravity, for example the string theory, we investigate a polynomial $f(R )$ inflation model in detail. We calculate the spectral index and tensor-to-scalar ratio in the $f(R )$ inflation model with the form of $f(R )=R+{R^2\over 6M^2}+{\lambda_n\over 2n}{R^n\over (3M^2)^{n-1}}$. Compared to Planck 2013, we find that $R^n$ term should be exponentially suppressed, i.e. $|\lambda_n|\lesssim 10^{-2n+2.6}$.

\end{abstract}


\maketitle


\section{Introduction}

Planck releases its first cosmological results \cite{Ade:2013zuv,Ade:2013uln}, and it strongly supports the six-parameter base $\Lambda$CDM model. In particular, Planck data prefers a highly significant deviation from scale-invariance of the primordial power spectrum, and the constraint on spectral index is given by 
\m
n_s=0.9603\pm 0.0073, 
\label{planckns}
\n
at $68\%$ CL. Actually it is almost the same as that from the pre-Planck data \cite{Cheng:2013iya}: 
\m
n_s=0.961\pm 0.007. 
\n
In addition, the constraint on the tensor-to-scalar ratio $r$ from Planck 2013 is 
\m
r<0.11
\label{planckr}
\n
at $95\%$ CL, which is also the same as that from the pre-Planck data in \cite{Cheng:2013iya}.

Recently a simple inflation model proposed by Starobinsky in 1980 \cite{Starobinsky:1980te} attracted attention of cosmologists. In this model, the inflationary expansion of the Universe is driven by the higher derivative terms in the action which takes the form 
\m
S={1\over 2\kappa^2}\int d^4x\sqrt{-g} \(R+{R^2\over 6M^2}\), 
\n
where $R$ is the Ricci scalar and $M$ is an energy scale. 
This model \cite{Mukhanov:1981xt,Starobinsky:1983zz} predicts that 
\m
n_s=1-{2\over N},\ r={12\over N^2},
\n
which are compatible with Planck 2013 nicely, where $N\simeq 50\sim 60$ is the number of e-folds before the end of inflation.
On the other hand, from the viewpoint of an ultraviolet(UV) complete quantum theory of gravity, for example the string theory, $\alpha'=1/M_s^{2}$ corrections to the Einstein-Hilbert action are always expected \cite{Polchinski:1998rq}, i.e. 
\m
S={1\over 2\kappa^2}\int d^4x\sqrt{-g} (R+c_2 \alpha' R^2+\sum_{i=3} c_i \alpha'^{i-1}R^i  \nonumber \\
+\hbox{other higher derivative terms}), 
\n
where $c_i$ are the dimensionless couplings. The higher derivative terms may also originate from the supergravity  \cite{Farakos:2013cqa,Ferrara:2013kca}. Here one point we want to mention is that in \cite{Ferrara:2013kca} the correction term of $R^4$ was proposed to be suppressed by the Planck scale $M_p=1/\kappa$, not $M_s$, in the new minimal supergravity model. More relevant fundamental aspects on the gravity with higher derivatives are considered in \cite{Benedetti:2012dx,Dietz:2012ic,Briscese:2013lna} etc.

In this paper we will phenomenologically investigate the inflation model with a polynomial $f(R )=R+{R^2\over 6M^2}+{\lambda_n\over 2n}{R^n\over (3M^2)^{n-1}}$ in detail, and we find that the slow-roll inflation can be achieved as long as the dimensionless coupling $\lambda_n$ is much smaller than one. 
This paper is organized as follows. In Sec.~2 a general discussion for the $f(R )$ inflation model is presented. We focus on the polynomial $f(R )$ inflation model in Sec.~3. Discussion and conclusion will be contained in Sec.~4.

\section{General $f(R )$ inflation model}

Let's start with the general $f(R )$ gravity in which the action takes the form  
\m
S={1\over 2\kappa^2}\int d^4x\sqrt{-g}f(R ), 
\n
where 
\m
f(R )=R+F(R ), 
\n
and $\kappa^2=8\pi G_N$.
If $F_{,RR}\neq 0$, the above action is equivalent to 
\m
S=\int d^4 x\sqrt{-g} \[{1\over 2\kappa^2}\varphi R - U(\varphi) \], 
\n
where 
\m
\varphi\equiv 1+F_{,\chi} (\chi), 
\label{vphichi}
\n
and 
\m
U(\varphi)={(\varphi-1) \chi(\varphi)-F(\chi(\varphi))\over 2\kappa^2}.
\n
We can do a conformal transformation to Einstein frame whose metric $g_{\mu\nu}^E$ is related to $g_{\mu\nu}$ by 
\m
g_{\mu\nu}^E=\varphi g_{\mu\nu}, 
\n
and the action in the Einstein frame becomes 
\m
S_E=\int d^4 x \sqrt{-g_E}\[{1\over 2\kappa^2}R_E-\half g_E^{\mu\nu} \p_\mu\phi \p_\nu\phi-V(\phi)\], 
\label{actionE}
\n
where 
\m
V(\phi)={1\over 2\kappa^2\varphi^2} \[(\varphi-1) \chi(\varphi)-F(\chi(\varphi))\],
\n
and $\varphi$ is related to the canonical field $\phi$ by 
\m
\varphi=e^{\sqrt{2/ 3}\kappa \phi}. 
\n
From now on, we will work in the Einstein frame and the subscript ``E" will be omitted. See a nice review in \cite{DeFelice:2010aj,Nojiri:2010wj}.

For the inflation govern by the action (\ref{actionE}), the slow-roll parameters $\epsilon$ and $\eta$ are respectively given by   
\m
\epsilon\equiv {1\over 2\kappa^2} \({V_\phi'\over V}\)^2={\varphi^2\over 3}\({V_\varphi'\over V}\)^2, 
\n
and 
\m
\eta\equiv {1\over \kappa^2} {V_\phi''\over V}={2\over 3}{\varphi V_\varphi'+\varphi^2 V_\varphi''\over V}. 
\n
The number of e-folds before the end of inflation is related to the value of $\varphi_N$ by 
\m
N\simeq \int_{t_N}^{t_{\rm end}}Hdt={3\over 2}\int_{\varphi_{\rm end}}^{\varphi_N} {V\over V_\varphi'} {d\varphi \over \varphi^2}. 
\n
The amplitude of the primordial scalar power spectrum takes the form 
\m
\Delta_{\cal R}^2={\kappa^4 V\over 24\pi^2\epsilon}, 
\label{deltaR2}
\n
and the spectral index $n_s$ and the tensor-to-scalar ratio $r$ are the the standard ones for the slow-roll inflation:  
\m
n_s&=&1-6\epsilon+2\eta,\\
r&=&16\epsilon. 
\n
These formula will be very useful for the following discussion.



\section{The polynomial $f(R )$ inflation model}

In this section we consider a polynomial $f(R )$ inflation model in which  
\m
f(R )=R+{R^2\over 6M^2}+{\lambda_n\over 2n} {R^n\over (3M^2)^{n-1}}, 
\label{pfr}
\n
where $\lambda_n$ is a dimensionless coupling and $n>2$. Here the extra coefficient of $R^2$ term is normalized to one by re-defining the energy scale $M$. It reduces to the Starobinsky's model when $\lambda_n\rightarrow 0$, and it approaches to the model with $f(R )=R+R^n/(3{\tilde M}^2)^{n-1}$ in the limit of $\lambda_n\gg 1$ which has been ruled out. See \cite{Kofman:1985aw,Maeda:1988ab,Kaneda:2010qv,Ketov:2010qz} and \cite{Martin:2013tda} for a review. The full dynamics for $n=4$ is analyzed in \cite{Saidov:2010wx} in detail. In this paper we focus on another limit, namely $\lambda_n\ll 1$ and the term of $R^n$ can be taken as a small correction to the Starobinsky's inflation model. In this limit the model can be expanded around the Starobinsky's model, and we do not need to do the full analyses of the dynamics.

\begin{figure}[ht]
\centerline{\includegraphics[width=\figurewidth]{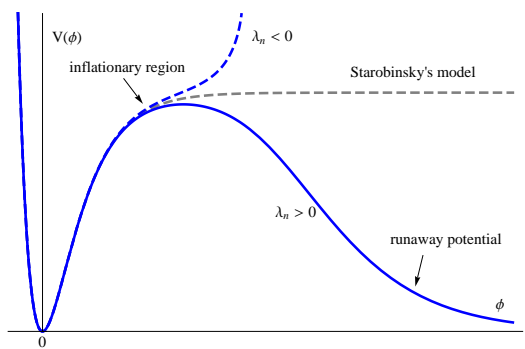}}
\caption{The potential of $\phi$. }
\label{fig:potential}
\end{figure}

From Eq.~(\ref{vphichi}), we obtain 
\m
{\lambda_n\over 2}\({\chi\over 3M^2}\)^{n-1}+{\chi\over 3M^2}-(\varphi-1)=0.
\label{vphichin}
\n
If $\lambda_n\neq 0$, in principle there are $(n-1)$ solutions for the above equation if $n$ is an integer. Here we only consider the solution which can reduce to that in the Starobinsky's model in the limit of $\lambda_n\rightarrow 0$. For example, for $n=3$, the solution of Eq.~(\ref{vphichin}) is given by 
\m
\chi(\varphi)={3M^2\over \lambda_3} \(\sqrt{1+2\lambda_3(\varphi-1)}-1\). 
\label{chivarphi3}
\n
For an arbitrary power $n$, we cannot find the analytic solution. Here we will figure out a general discussion for the arbitrary power $n$ approximately. 
The potential of $\phi$ is illustrated in Fig.~\ref{fig:potential}. 

First of all, let's see the asymptotic behaviour of $V(\phi)$. For $\lambda_n\varphi^{n-2}\gg 1$, the solution of Eq.~(\ref{vphichin}) is roughly given by 
\m
\chi(\varphi)\simeq 3M^2 \({2\varphi\over \lambda_n}\)^{1\over n-1}. 
\n
Therefore the potential of $\phi$ reads 
\m
V(\phi)\simeq {3(n-1)M^2\over 2n \kappa^2}\({2\over \lambda_n}\)^{1\over n-1} e^{-{n-2\over n-1}\sqrt{2\over 3}\kappa \phi}, 
\n
which is nothing but a runaway potential. If the dynamics of inflation is govern by this potential, the slow-roll parameters are 
\m
\epsilon&=&{1\over 3}\({n-2\over n-1}\)^2, \\
\eta&=& {2\over 3}\({n-2\over n-1}\)^2, 
\n
and then the spectral and the tensor-to-scalar ratio become 
\m
n_s&=&1-{2\over 3}\({n-2\over n-1}\)^2,\\
r&=&{16\over 3}\({n-2\over n-1}\)^2. 
\n
Considering that $n$ is an integer and not less than 3, we find that $n_s\in [1/3,5/6]$ and $r\in [4/3,16/3]$. It indicates that the inflation driven by the runaway potential in the asymptotic region has been ruled out by the observations unless $n$ is fine-tuned to close to two ($n=2.32$ for $n_s=0.96$). Such a value of $n$ is quite unnatural from the viewpoint of a fundamental theory.

Now let's move to another region of the potential where $\lambda(\varphi-1)^{n-2} \ll 1$. In this limit  the approximate solution of Eq.~(\ref{vphichin}) is 
\m
\chi(\varphi)\simeq 3M^2 (\varphi-1) \[1-{\lambda_n\over 2}(\varphi-1)^{n-2}\], 
\n
and thus the potential of $\varphi$ becomes 
\m
V(\varphi)\simeq {3M^2\over 4\kappa^2}(1-\varphi^{-1})^2 \[1-{\lambda_n\over n}(\varphi-1)^{n-2}\]. 
\n
During inflation $\varphi\gg 1$ and then the potential can be simplified to be 
\m
V(\varphi)\simeq {3M^2\over 4\kappa^2} \(1-2\varphi^{-1}-{1\over n}\lambda_n \varphi^{n-2}\). 
\label{vvar}
\n
Therefore the slow-roll parameters can be written down by   
\m
\epsilon&\simeq&{4\over 3} \(\varphi^{-1}-{n-2\over 2n}\lambda_n \varphi^{n-2}\)^2, \label{depsilon}\\
\eta&\simeq& -{4\over 3}\[\varphi^{-1}+{(n-2)^2\over 2n}\lambda_n \varphi^{n-2}\]. \label{deta}
\n
\noindent $\bullet$ If $\lambda_n>0$, there is a top of potential located at $\varphi=\varphi_t$ where 
\m
\varphi_t=\({2n\over n-2}{1\over \lambda_n}\)^{1\over n-1}, 
\n
which divides the potential into two regions: $\varphi<\varphi_t$ and $\varphi>\varphi_t$. However the region of $\varphi>\varphi_t$ corresponds to large $R$ and there is no graceful exit for inflation. Therefore we only focus on the region of $\varphi<\varphi_t$ where the inflation can occur and naturally end when the inflaton field oscillates around its local minimum at $\phi=0$.  \\
\noindent $\bullet$ If $\lambda_n<0$, there is no top of the potential. See the dashed curve in Fig.~\ref{fig:potential}. In \cite{Appleby:2009uf} the classical and quantum stability of $f(R )$ model requires that 
\m
f'(R)>0,\quad \hbox{and}\quad f''(R)>0. 
\n
These two inequalities are satisfied for $\lambda_n>0$. However for $\lambda_n<0$, one needs to worry about the condition of $f''(R)>0$. \footnote{The condition of $f'(R)=\varphi=0$ corresponds to $\phi\rightarrow -\infty$ where the inflaton field never reaches. Thus we do not need to worry about the condition of $f'(R)>0$ in this case. } The turning point from positive $f''(R)$ to negative one happens when $f''(R)=0$, namely 
\m
{R\over 3M^2}=\({-2\over (n-1)\lambda_n}\)^{1\over n-2}. 
\n 
The requirement of $f''(R)>0$ implies an upper bound on the field $\varphi$, namely
\m
\varphi-1<{n-2\over n-1}\({-2\over (n-1)\lambda_n}\)^{1\over n-2}. 
\label{ubvarphi}
\n
For $n=3$, $\varphi-1< -{1/ (2\lambda_3)}$ which is the same as the requirement that the term in the square root of Eq.~(\ref{chivarphi3}) be non-negative. This upper bound on $\varphi$ implies that this case is UV-incomplete. In this paper we focus on $|\lambda_n|\ll 1$, and inflation happens when the inequality (\ref{ubvarphi}) is satisfied.


The number of e-folds before the end of inflation is related to value of $\varphi_N$ by 
\m
N&\simeq& {3\over 4} \int_{\varphi_{\rm end}}^{\varphi_N} {1\over 1-{n-2\over 2n}\lambda_n \varphi^{n-1}} d\varphi \nonumber \\
&\simeq& {3\over 4} \varphi_N\ _2F_1\({1\over n-1},1,{n\over n-1},{n-2\over 2n}\lambda_n \varphi_N^{n-1}\), 
\label{regionIN}
\n
where $_2F_1$ is the hypergeometric function. For $\lambda_n=0$, $_2F_1({1\over n-1},1,{n\over n-1},0)=1$ and then $\varphi_N={4N/3}$. Here we consider the term of $\lambda_n \varphi_N^{n-1}$ is much smaller than one and then Eq.~(\ref{regionIN}) becomes 
\m
N\simeq {3\over 4}\varphi_N\(1+{n-2\over 2n^2}\lambda_n \varphi_N^{n-1}\), 
\n
or equivalently, 
\m
\varphi_N\simeq {4\over 3}N\[1-{n-2\over 2n^2}\lambda_n \({4\over 3}N\)^{n-1}\]. 
\n
Substituting the above result into Eqs.~(\ref{depsilon}) and (\ref{deta}),  we obtain  
\m
\epsilon&\simeq& {3\over 4N^2} \[1-{(n-1)(n-2)\over n^2}\lambda_n \({4\over 3}N\)^{n-1}\],\\
\eta&\simeq&-{1\over N}-{2(n-1)^2(n-2)\over 3n^2}\lambda_n \({4\over 3}N\)^{n-2}.
\n

\begin{figure}[ht]
\centerline{\includegraphics[width=\figurewidth]{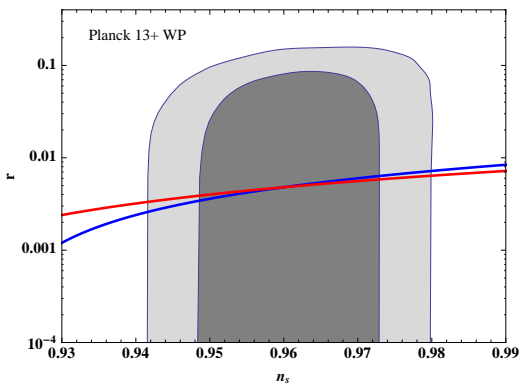}}
\caption{The plot of $r-n_s$ for the polynomial $f(R )$ inflation model. The blue and red curves respectively correspond to $n=3$ and $n=4$ by varying $\lambda_n$. Here the gray and light gray regions correspond to the constraints from Planck 13 and WMAP polarization (WP) data at $68\%$ and $95\%$ CL respectively. }
\label{fig:rnsI}
\end{figure}

Using Eq.~(\ref{deltaR2}), the amplitude of the primordial scalar power spectrum becomes 
\m
\Delta_{\cal R}^2\simeq {\kappa^2M^2N^2\over 24\pi^2}. 
\n
Planck normalization \cite{Ade:2013zuv,Ade:2013uln} is $\Delta_{\cal R}^2=2.2\times 10^{-9}$ which implies $M/M_p\simeq 1.44\times 10^{-5}$ for $N=50$. The spectral index and tensor-to-scalar ratio are respectively given by 
\m
n_s\simeq 1-{2\over N}-{4(n-1)^2(n-2)\over 3n^2} \lambda_n \({4\over 3}N\)^{n-2},  
\n
and 
\m
r={12\over N^2} \[1-{(n-1)(n-2)\over n^2}\lambda_n \({4\over 3}N\)^{n-1}\]. 
\n
Actually the above formula are also applicable for the case of $\lambda_n<0$, and the tensor-to-scalar ratio can be a little bit enhanced for $\lambda_n<0$ if $n_s>0.96$. See the plot of $r-n_s$ in Fig.~\ref{fig:rnsI}. 
Adopting the numerical method, we find that the dimensionless coupling should satisfy 
\m
|\lambda_n|\lesssim 10^{-2n+2.6}, 
\n
which is much smaller than one. Otherwise, the prediction of the polynomial $f(R )$ inflation conflicts with Planck 2013 at more than $95\%$ CL. In addition, one can also calculate the running of spectral index, namely 
\m
{dn_s\over d\ln k}\simeq -{2\over N^2}+{16(n-1)^2(n-2)^2\over 9n^2}\lambda_n \({4\over 3}N\)^{n-3}, 
\n
which is order of $-2/N^2=-8\times 10^{-4}$. It shows that the running of spectral index in this model is too small to be detected by the observations in the near future.

\section{Discussion}

Even though the Starobinsky's inflation model can fit the Planck data very well, the higher derivative corrections to the Starobinsky's model is quite generic from the viewpoint of the UV complete quantum gravity theory, such as string theory and supergravity. In this paper, as a toy model, we carefully investigate the polynomial $f(R )$ inflation model whose form is taken in Eq.~(\ref{pfr}), and we find that the slow-roll inflation can happen as long as the dimensionless coupling $\lambda_n$ is small enough. The smallness of the dimensionless coupling constant $\lambda_n$ does not certainly imply that a fine-tuning is necessary. For example, in \cite{Ferrara:2013kca}, $R^4$ term was proposed to be suppressed by the Planck scale and then the upper bound on the effective dimensionless coupling becomes $\xi\sim \lambda_4 (M_p/M)^6\lesssim 10^{26}$ which is much larger than one.

In this paper we only focus on the model with a $R^n$ correction term to the Starobinsky's inflation model.  Actually some other correction terms, such as $R_{\mu\nu}R^{\mu\nu}$, $R_{\mu\nu\rho\sigma}R^{\mu\nu\rho\sigma}$ and so on, are also expected in the UV complete quantum theory of gravity. A well-motivated model is the so-called asymptotically safe inflation \cite{Weinberg:2009wa,Tye:2010an,Fang:2012ca}. Unfortunately, such a model suffers from a serious ghost instability and cannot be self-consistent \cite{Fang:2012ca}. Anyway, how to naturally build up an inflation model in a fundamental theory is still an open question.

\noindent {\bf Acknowledgments}

QGH would like to thank B.~Chen and P.~X.~Wu for useful conversation. 
This work is supported by the project of Knowledge Innovation Program of Chinese Academy of Science and grants from NSFC (grant NO. 10821504, 11322545 and 11335012).



\end{document}